\documentclass[11pt,a4paper]{article}
\pdfoutput=1 
\usepackage{jcappub}

\usepackage{ifpdf}
\usepackage{mathtools}
\usepackage{bm}
\usepackage{tikz}
\usepackage[tableposition=top]{caption}

\begin{document}


\title{Merger rate of initially clustered primordial black holes for the two-body channel}
\author[a,b]{Kentaro Kasai,}
\author[c,b]{Masahiro Kawasaki,}
\author[d,e]{Kai Murai,}
\author[b]{Shunsuke Neda}
\affiliation[a]{Theory Center, Institute of Particle and Nuclear Studies (IPNS), High Energy Accelerator Research Organization (KEK), 1-1 Oho, Tsukuba, Ibaraki 305-0801, Japan}
\affiliation[b]{ICRR, University of Tokyo, Kashiwa 277-8582, Japan}
\affiliation[c]{Kavli IPMU (WPI), UTIAS, University of Tokyo, Kashiwa 277-8583, Japan}
\affiliation[d]{RIKEN Center for Interdisciplinary Theoretical and Mathematical Sciences (iTHEMS), RIKEN, Wako 351-0198, Japan}
\affiliation[e]{Department of Physics, Tohoku University, Sendai 980-8578, Japan}

\abstract{%
Primordial black holes (PBHs) may form an initially clustered population depending on their production mechanism. 
Motivated by binary black-hole merger events observed by gravitational-wave interferometers, we revisit the evaluation of the merger rate of PBH binaries and extend the formalism to include the effects of clustering.
We show that, in the presence of relatively weak PBH clustering, the LIGO-Virgo-KAGRA events can be explained with a smaller value of $f_{\rm PBH}$ than in scenarios with Poisson-distributed PBHs, at least in the early two-body channel.
However, for stronger clustering, the merger rate in the two-body channel is significantly suppressed due to the formation of three-body systems.
}


\emailAdd{kkasai98@post.kek.jp}
\emailAdd{masahiro.kawasaki@ipmu.jp}
\emailAdd{kai.murai@riken.jp}
\emailAdd{neda@icrr.u-tokyo.ac.jp}

\begin{flushright}
    TU-1280, RIKEN-iTHEMS-Report-26
\end{flushright}

\maketitle

\section{Introduction}
\label{sec: intro}

Primordial black holes (PBHs)~\cite{Zeldovich:1967lct,Hawking:1971ei,Carr:1974nx,Carr:1975qj} are a possible observable probe of the early universe before the Big Bang Nucleosynthesis epoch.
Recent developments in gravitational wave observations enable us to detect binary black hole (BH) mergers.
The LIGO-Virgo-KAGRA (LVK) collaboration reports the inferred merger rate of binary black holes as $17.9\,\text{--}\,44\,\mathrm{Gpc}^{-3} \mathrm{yr}^{-1}$~\cite{KAGRA:2021duu}.
This result implies the PBH fraction to dark matter of $f_\mathrm{PBH} \sim 5\times10^{-3}$~\cite{Raidal:2024bmm} under the assumptions of a uniform spatial distribution and a monochromatic mass function~\cite{Nakamura:1997sm}.

Following the pioneering work decades ago~\cite{Nakamura:1997sm,Ioka:1998nz}, many studies have investigated the merger rate of PBHs~\cite{Sasaki:2016jop,Raidal:2017mfl,Ali-Haimoud:2017rtz,Kocsis:2017yty,Chen:2018czv,Kavanagh:2018ggo,Ballesteros:2018swv,Belotsky:2018wph,Bringmann:2018mxj,Raidal:2018bbj,Liu:2018ess,Liu:2019rnx,Ding:2019tjk,Garriga:2019vqu,Vaskonen:2019jpv,Young:2019gfc,Gow:2019pok,Wu:2020drm,Atal:2020igj,DeLuca:2020jug,Kawasaki:2021zir,Deng:2021gkx,Stasenko:2021wej,Okano:2022nkb,Franciolini:2022ewd,Fakhry:2022zum,Fakhry:2022hzh,Tkachev:2022alb,Jangra:2023mqp,Stasenko:2023zmf,Fakhry:2023fbt,Hai-LongHuang:2023atg,Ding:2023smy,Raidal:2024bmm,Stasenko:2024pzd,Aljaf:2024fru,Ding:2024mro,Aljaf:2025dta,Stasenko:2025vqz}, motivated in part by the first gravitational-wave event observed by LIGO~\cite{LIGOScientific:2016aoc}.
Although many studies assume uniform PBH distributions in evaluating the merger rate, clustering is important from both theoretical and observational perspectives.
Theoretically, PBHs are formed from the collapse of large density fluctuations in the early universe.
If such density fluctuations have spatial correlations, the resulting PBHs are clustered.
In fact, strong clustering of PBHs is predicted in some concrete models for PBH formation \cite{Nakama:2016kfq,Hasegawa:2017jtk,Hasegawa:2018yuy,Kawasaki:2019iis,Kitajima:2020kig,Kawasaki:2021zir,Kasai:2022vhq,Kasai:2023ofh,Kasai:2023qic} (see also Refs.~\cite{Shinohara:2021psq,Shinohara:2023wjd} for observational analyses).

Observationally, PBH clustering affects the present-day merger rate of binary PBHs.
If PBHs are clustered, the probability distribution of the initial separation of PBHs is modified, which further affects the distribution of PBH binaries with respect to the semi-major axis and angular momentum.
For instance, the PBH clustering can be represented as an enhancement of the PBH number density on certain scales or in local regions~\cite{Raidal:2017mfl,Bringmann:2018mxj,Ding:2019tjk,Vaskonen:2019jpv,Young:2019gfc,Stasenko:2025vqz}.
More precisely, the effects of initial clustering of PBHs can be encoded into the two-point correlation function of the PBH number density, which affects the formation rate of PBH binaries~\cite{Ballesteros:2018swv,Atal:2020igj,Kawasaki:2021zir}.
While these studies assume that the angular momentum of PBH binaries is induced only by the nearest external PBH, the contributions to the angular momentum from more distant PBHs and dark matter fluctuations are important for a precise estimation of the merger rate of PBHs~\cite{Ali-Haimoud:2017rtz} (see also Refs.~\cite{Raidal:2018bbj,Raidal:2024bmm}).

In this paper, we extend the previous method~\cite{Raidal:2018bbj,Raidal:2024bmm} to evaluate the merger rate for two-body channel by including the effect of PBH clustering.
Roughly speaking, the merger rate of PBH binaries can be calculated from the probability distribution of binary formation rate with respect to the semi-major axis and angular momentum of binaries at formation (see Eq.~\eqref{eq: merger rate original} below).
Since the merger time is determined by the semi-major axis and angular momentum, we finally obtain the event rate of mergers observed at the current time by integrating the probability distribution of binary formation under the condition that the coalescence time equals the age of the universe.
Here, the initial clustering of PBHs can affect both the semi-major axis and angular momentum.
We first revisit the formation of PBH binaries and then derive the probability distribution of binary formation with respect to the semi-major axis.
Then, we review the estimation of the angular momentum of black hole binaries and expand the formulation so that we can include the effect of PBH clustering.

Further, we need to consider the disruption of binaries in dense environments such as halos.
PBH binaries can be disrupted by accidental encounters with other PBHs and by the core collapse of the host halo.
We incorporate the effects of PBH clustering on halo formation and derive the suppression factor of the merger rate due to binary disruption.

Finally, we evaluate the merger rate for initially clustered PBHs.
We find that the merger rate inferred from the LVK observations can be accounted for with and without PBH clustering.
The favored value of the PBH fraction in dark matter, $f_\mathrm{PBH}$, is smaller with clustering.
However, if the clustering is too strong, the merger rate is significantly suppressed for large $f_\mathrm{PBH}$, and the inferred merger rate cannot be obtained.
This is due to the enhanced formation of three-body systems, in agreement with previous work~\cite{Kawasaki:2021zir}.
In addition, we find that the suppression due to binary disruption becomes less important for stronger clustering.
With PBH clustering, the typical halo size is larger, which reduces the probability of binary disruption relative to the case without clustering.

The rest of this paper is organized as follows.
In Sec.~\ref{sec: rev}, we overview the evaluation of the merger rate for Poisson-distributed PBHs.
In Sec.~\ref{sec: est}, we continue reviewing the angular momentum estimate of PBH binaries and extend the formalism to include the PBH clustering.
We argue the suppression factor for the merger rate due to the astrophysical effects in Sec.~\ref{sec: suppression}.
In Sec.~\ref{sec: res}, we show the resulting merger rate.
Finally, we conclude in Sec.~\ref{sec: disc}.

\section{Overview of the PBH merger rate evaluation}
\label{sec: rev}

\begin{figure}
\centering
\begin{tikzpicture}[scale=0.5]
  \coordinate (P) at ({3*cos(240)}, {3*sin(240)});
  \coordinate (Q) at ({1*cos(150)}, {1*sin(150)});
  \coordinate (R) at ({-1*cos(150)}, {-1*sin(150)});
  \draw[thick,gray] (0,0) circle (3cm);
  \draw[<->, gray] (0,0) -- (P);
  \draw[-] (Q) -- (R);
  \fill (Q) circle (5pt);
  \fill (R) circle (5pt);
  \fill (-1.5,3.5) circle (5pt);
  \fill (1.5,-3) circle (5pt);
  \fill (-4,-2) circle (5pt);
  \fill (-3,2) circle (5pt);
  \fill (2.5,2.5) circle (5pt);
  \fill (3.5,-1.5) circle (5pt);
  \node[gray] at ({1.5*cos(260)}, {1.5*sin(260)}) {$y$};
  \node at ({0.45*cos(60)}, {0.45*sin(60)}) {$x_\mathrm{in}$};
  \node at (0, -4.2) {\text{1.\, Binary~formation}};

  \draw[gray] (10,0) circle [x radius = 3.7, y radius = 1.2];
  \fill (12.4,0.9) circle (10pt);
  \fill (7.6,-0.9) circle (10pt);
  \draw (12.4-0.4,0.9+1.1) arc [start angle = 100, end angle = 0, radius = 1.3];
  \draw (12.4-0.5,0.9+1.6) arc [start angle = 100, end angle = 0, radius = 1.8];
  \draw (12.4-0.6,0.9+2.1) arc [start angle = 100, end angle = 0, radius = 2.3];
  \node at (10, -4.2) {\text{2.\, Binary~shrinking}};
  \node at (12.4+2.6, 0.9+1.6) {\text{GW}};

  \fill (20.3,0.1125) circle (10pt);
  \fill (19.7,-0.1125) circle (10pt);
  \draw[ultra thick] (20.3-0.4,0.1125+1.1) arc [start angle = 100, end angle = 0, radius = 1.3];
  \draw[ultra thick] (20.3-0.5,0.1125+1.6) arc [start angle = 100, end angle = 0, radius = 1.8];
  \draw[ultra thick] (20.3-0.6,0.1125+2.1) arc [start angle = 100, end angle = 0, radius = 2.3];
  \node at (20.3+2.6, 0.1125+1.6) {\text{GW}};
  \node at (20, -4.2) {\text{3.\, Binary merger}};
\end{tikzpicture}
\caption{%
    Early two-body merger channel of PBHs.
    First, two PBHs form a binary with a sufficiently small initial separation $x_\mathrm{in}$.
    Here, we assume that no other PBHs exist within the distance of $y$ from the binary to avoid the formation of a three-body system.
    Once a binary is formed, it receives angular momentum from external PBHs and matter fluctuations and rotates in an oval orbit.
    The binary loses its energy and angular momentum by emitting gravitational waves, and the binary orbit shrinks.
    Finally, the binary merges and emits the specific signal of gravitational waves.
    }
\label{fig: early 2b}
\end{figure}
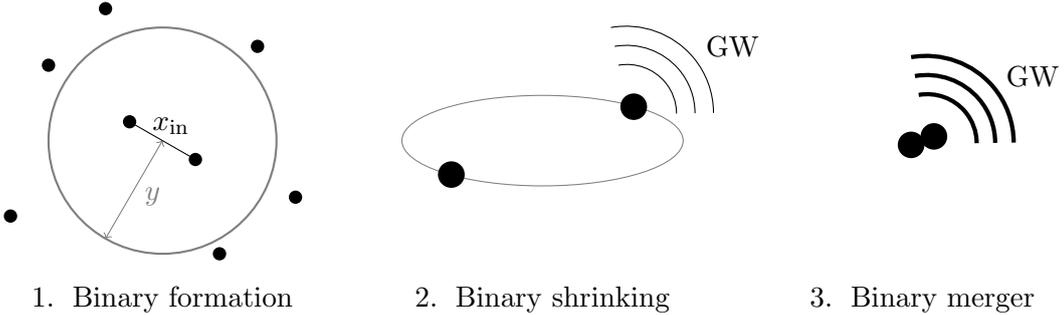

Here, we overview the evaluation of the PBH merger rate following Ref.~\cite{Raidal:2024bmm}.
While Ref.~\cite{Raidal:2024bmm} studies mergers of PBHs spatially distributed according to the Poisson distribution, the formalism presented in this section can also be applied to initially clustered PBHs, which we will discuss in the following sections.
In this paper, we focus on the early two-body merger channel, which corresponds to an isolated binary black hole merger shown in Fig.~\ref{fig: early 2b}.
To this end, we consider cases where no other PBHs exist within a certain distance $y$ from the binary.
Otherwise, the binary together with its nearest PBH neighbor would form a three-body system.
Mergers through three-body systems, i.e., the early three-body merger channel, should be treated separately, which will be addressed in future work (see, e.g., Refs.~\cite{Franciolini:2022ewd,Raidal:2024bmm} for some discussion of the three-body channel).
After formation, the binary loses its energy and angular momentum through gravitational wave emission and eventually merges.
In the following, we explain the PBH merger process step by step.

\subsection{PBH binary}
\label{subsec: dec}

At the beginning of the PBH merger process, two PBHs form a binary.
The formation of a binary can be roughly estimated by comparing the free fall and the cosmic expansion~\cite{Nakamura:1997sm}.
For two PBHs with the comoving separation $x_\mathrm{in}$, the typical free-fall timescale is given by
\begin{align}
    t_\mathrm{ff}
    =
    \frac{(a x_\mathrm{in})^\frac{3}{2}}{\sqrt{2G(M/2-\bar{\rho}_\mathrm{PBH} a^3 V(x_\mathrm{in}))}} ,
    \label{eq: free-fall}
\end{align}
where $a$ is the scale factor, $G$ is the Newton constant, $M$ is the total mass of the binary system, $V(x_\mathrm{in})\equiv4 \pi x_\mathrm{in}^3/3$ is the comoving volume with comoving radius $x_\mathrm{in}$, and $\bar{\rho}_{\rm{PBH}}$ is the average energy density of PBHs.
Note that $\bar{\rho}_\mathrm{PBH}$ scales as $\propto a^{-3}$, and the comoving density $\bar{\rho}_\mathrm{PBH} a^3$ in the denominator is constant in time.
Here, we generalize the original free-fall time in Newtonian dynamics to a relativity-like form by subtracting the spatially averaged mass in radius $x_\mathrm{in}$ from $M/2$.
On the other hand, the typical timescale of the cosmic expansion, i.e., the Hubble time $t_H$, is given by
\begin{align}
    t_H
    \equiv 
    H^{-1}
    =
    \left[\frac{8 \pi G}{3} \bar{\rho}_{\mathrm{m},\mathrm{eq}} \left(\frac{a_\mathrm{eq}}{a}\right)^{4}\right]^{-\frac{1}{2}} ,
\end{align}
where $\bar{\rho}_\mathrm{m}$ is the energy density of the matter component, including baryons, PBHs, and other dark matter components, and the subscript ``eq'' denotes the quantity at the matter-radiation equality.
Here, we assumed the radiation-dominated universe.
If the free-fall time is shorter than the Hubble time, the two PBHs decouple from the Hubble flow and form a binary.
Since $t_H \,(\propto a^2)$ increases faster than $t_\mathrm{ff} \,(\propto a^{3/2})$, we can consider that a binary is formed when $t_H \sim t_\mathrm{ff}$.
This condition gives the scale factor at the decoupling $a_\mathrm{dc}$ as
\begin{align}
    a_\mathrm{dc}
    =
    \frac{a_\mathrm{eq}}{\delta_\mathrm{pair}} ,
    \label{eq: a_dc}
\end{align}
where $\delta_\mathrm{pair}$ is the overdensity given by
\begin{align}
    \delta_\mathrm{pair}
    =
    \frac{M/2-\bar{\rho}_{\mathrm{PBH},0}V(x_\mathrm{in})}{\bar{\rho}_{\mathrm{m},0}V(x_\mathrm{in})} .
    \label{eq: delta_pair}
\end{align}
Here, the quantities accompanied by the subscript ``0'' denote their present values, and we normalize the scale factor by $a_0 = 1$.
Note that $t_H$ increases as $\propto a^{3/2} \propto t_\mathrm{ff}$ in the matter-dominated universe.
Thus, a pair of PBHs that does not form a binary in the radiation-dominated era never forms a binary in the matter-dominated era.

\begin{figure}[t]
    \centering
    \includegraphics[width=.75\textwidth ]{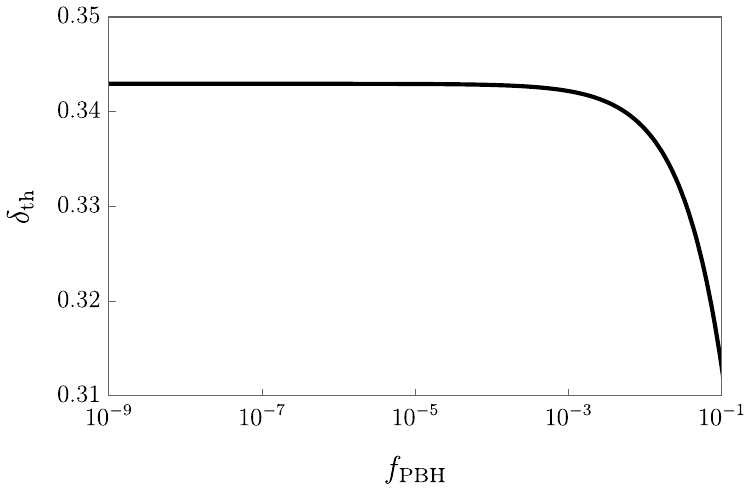}
    \caption{
        Threshold for binary formation in the initial density contrast $\delta_\mathrm{pair}$.
    }
    \label{fig: bin form}
\end{figure}

To clarify the relation between the Hubble flow and binary formation more precisely, we solve the equation of motion for PBHs numerically.
Assuming the ellipticity $e$ of the binary orbit is nearly equal to unity,%
\footnote{This assumption of $e \simeq 1$ is justified for PBH binaries whose merger is observed at the current time \cite{Raidal:2024bmm}.}
we can consider that PBHs move in one-dimensional space in the radial direction.
Then, the equation of motion for the PBH separation is given by
\begin{align}
    \ddot{r}-\frac{\ddot{a}}{a}r+G\frac{M-2\bar{\rho}_{\mathrm{PBH}} a^3 V(|r|/a)}{r^{2}}\frac{r}{\lvert r\rvert}
    =
    0 ,
\end{align}
where $r$ is the coordinate denoting the physical distance between the two PBHs, and the dot denotes the derivative with respect to the cosmic time $t$.
Note that we allow negative values of $r$, which represent the opposite relative position of PBHs compared to positive $r$.
This equation is numerically solved following the procedure in Ref.~\cite{Raidal:2024bmm}.
For the comoving coordinate $x=r/a$, the equation of motion becomes
\begin{align}
    \ddot{x}+2\frac{\dot{a}}{a}\dot{x}
    +
    G\frac{M-2\bar{\rho}_{\mathrm{PBH}} a^3 V(x)}{a^3x^2}\frac{x}{\lvert x\rvert}
    =
    0 .
    \label{eq: binary_formation_0}
\end{align}
We set the initial condition at $a=a_i$, when $x$ remains effectively constant due to the Hubble friction, as
\begin{align}
    x=x_\mathrm{in},\quad \frac{\mathrm{d}x}{\mathrm{d}t}=0 .
\end{align}
Since $H\equiv\dot{a}/a=1/(2t)$ during the radiation-dominated era, Eq.~\eqref{eq: binary_formation_0} is written as
\begin{align}
    \left(\frac{\mathrm{d}}{\mathrm{d} \ln{a}}\right)^2x
    +
    \frac{a}{a_\mathrm{dc}}\frac{x_\mathrm{in}^3}{x^2}
    \frac{M-2\bar{\rho}_{\mathrm{PBH},0}V(x)}{M-2\bar{\rho}_{\mathrm{PBH},0}V(x_\mathrm{in})}\frac{x}{\lvert x\rvert}
    =
    0 .
\end{align}
Introducing $\chi(\tilde{a}) \equiv x/x_\mathrm{in}$ and $\tilde{a} \equiv a/a_\mathrm{eq}$, we obtain
\begin{align}
    \label{eq:binary_formation}
    \left(\tilde{a}\frac{\mathrm{d}}{\mathrm{d}\tilde{a}}\right)^2\chi+\frac{\tilde{a}}{\chi^2}\frac{\chi}{\lvert \chi\rvert}
    \left(\delta_\mathrm{pair}+ \frac{\Omega_\mathrm{DM}}{\Omega_\mathrm{m}} f_\mathrm{PBH}(1-\chi^3)\right)
    =
    0
    ,
\end{align}
with the initial condition of
\begin{align}
    \chi(a_i/a_\mathrm{eq})=1,
    \quad
    \frac{\mathrm{d}\chi}{\mathrm{d}\tilde{a}}
    (a_i/a_\mathrm{eq})=0.
\end{align}
Here, $\Omega_\mathrm{m}$ and $\Omega_\mathrm{DM}$ are the present density fractions of matter and dark matter, respectively. 
With this equation, we can obtain the evolution of $\chi$ for given $\delta_\mathrm{pair}$ and $f_\mathrm{PBH}$.
By solving Eq.~\eqref{eq:binary_formation} numerically, it is found that $\tilde{a}\chi$ oscillates with the amplitude $\tilde{a}\chi_\mathrm{amp}$ fitted by
\begin{align}
    \tilde{a}\,\chi_\mathrm{amp}(\tilde{a})
    \simeq
    \frac{0.209}{\delta_\mathrm{pair}\left[1+0.230 \left(\frac{f_\mathrm{PBH}}{\delta_\mathrm{pair}}\right)^{0.822}\right]}.
    \label{eq: binary_amp}
\end{align}
The scale factor at the binary formation $a_\mathrm{form}$, which we define as the first time of $\dot{r}=0$, is also numerically obtained, and the fitting formula is given by 
\begin{align}
    \frac{a_\mathrm{form}}{a_\mathrm{eq}}
    \simeq
    \frac{0.343}{\delta_\mathrm{pair}\left[1+0.246 \left(\frac{f_\mathrm{PBH}}{\delta_\mathrm{pair}}\right)^{0.842}\right]}.
    \label{eq: binary_form}
\end{align}
For details of the fitting, see Appendix~\ref{app: valid}.
Since the binaries are formed only during the radiation-dominated era, we impose the binary formation condition of $a_\mathrm{form}/a_\mathrm{eq} < 1$ to obtain
\begin{align}
    \delta_\mathrm{pair} 
    \gtrsim
    \delta_\mathrm{th}
    \simeq
    0.343-0.187~f_\mathrm{PBH}^{0.795}.
    \label{eq: decoupling time}
\end{align}
We show this fitting formula in Fig.\,\ref{fig: bin form}.
This will put the upper limit on $x_\mathrm{in}$ in the evaluation of the merger rate since $\delta_\mathrm{pair}$ is a decreasing function of $x_\mathrm{in}$.%
\footnote{%
    While this condition for $\delta_\mathrm{pair}$ is also discussed in Ref.~\cite{Raidal:2024bmm}, the upper bound on $x_\mathrm{in}$ is not involved in evaluating the merger rate there.
    We find that the inclusion of this upper bound can considerably affect the resulting merger rate.
    See Appendix~\ref{app: wo clust consis check} for the comparison of the merger rates with and without the upper bound on $x_\mathrm{in}$.
}

Consequently, the semi-major axis is estimated as
\begin{align}
    r_a
    =
    \frac{a x_\mathrm{in}}{2}\chi_\mathrm{amp}
    \simeq
    \frac{0.8 \pi}{3}x_\mathrm{in}^4\frac{\bar{\rho}_{\mathrm{rad},0}}{M}
    \frac{1}{\left(1-\frac{8 \pi  x_\mathrm{in}^3 \bar{\rho} _{\mathrm{PBH},0}}{3
   M}\right)\left(1+0.230 \left(\frac{f_\mathrm{PBH}}{\delta_\mathrm{pair}}\right)^{0.822}\right)
   }.
    \label{eq: r_a}
\end{align}
Since Eq.\,\eqref{eq: decoupling time} restricts the last factor in Eq.~\eqref{eq: r_a} as
\begin{align}
    1\leq\frac{1}{\left(1-\frac{8 \pi  x_\mathrm{in}^3 \bar{\rho}_{\mathrm{PBH},0}}{3
   M}\right)\left(1+0.230 \left(\frac{f_\mathrm{PBH}}{\delta_\mathrm{pair}}\right)^{0.822}\right)
   }\lesssim1.16
   ,
    \label{eq: chi0 check}
\end{align}
for $f_\mathrm{PBH} < 0.1$, which will be our focus below, we neglect this factor for simplicity.

After the binary formation, the semi-major axis $r_a$ decreases through gravitational wave emission.
It is known that the coalescence time, assuming $e\simeq1$, is given by~\cite{Peters:1964zz}
\begin{align}
    \tau\sim\frac{3}{85}\frac{r_a^4\left|\bm{j}\right|^7}{G^3\mu M^2}
    ,
    \label{eq: coalescence}
\end{align}
where $\mu$ is the reduced mass of the binary system, and $\bm{j}$ is the dimensionless angular momentum of the binary defined below. 
Under the external gravitational potential $\phi_\mathrm{ex}$, the angular momentum is given by 
\begin{align}
    \bm{L}
    =
    - \mu \int \mathrm{d}t\,\bm{r}\times T\bm{r}
    ,
\end{align}
where $\bm{r}$ is the physical separation of the binary PBHs, and $T$ is a matrix whose components are defined by
\begin{align}
    T_{ij}
    =
    \partial_i \partial_j \phi_\mathrm{ex}(\bm{r}_c) ,
\end{align}
with $\bm{r}_c$ being the spatial coordinate of the center of mass of the binary.
Since $\phi_\mathrm{ex}$ scales as $\phi_\mathrm{ex}\propto a^{-1}$, $T$ follows $T \propto a^{-3}$~\cite{Raidal:2024bmm}.
The dimensionless angular momentum is defined by
\begin{align}
    \bm{j}
    &\equiv
    \frac{\bm{L}/\mu}{\sqrt{GMr_a}}
    =-
    \frac{1}{\sqrt{GMr_a}}\frac{1}{a^2 H}\hat{\bm{r}}\times(a^3T)\hat{\bm{r}}a_\mathrm{eq}x_\mathrm{in}^2\int_{0}^{\infty} \mathrm{d}\tilde{a} \, 
    \chi^2(\tilde{a})
    \nonumber \\
    &\simeq
    - 0.95\frac{x_\mathrm{in}^3}{GM}\hat{\bm{r}}\times a^3 T\hat{\bm{r}}.
    \label{eq: ang mom}
\end{align}
where $\hat{\bm{r}}\equiv \bm{r}/|\bm{r}|$.
Strictly speaking, the numerical coefficient here depends on $\delta_\mathrm{pair}$ and $f_\mathrm{PBH}$ via $r_a$ and $\chi(\tilde{a})$. 
However, we find that it hardly depends on these parameters, and thus we adopt the coefficient independent of $\delta_\mathrm{pair}$ and $f_\mathrm{PBH}$ presented in the previous study~\cite{Raidal:2024bmm}, which matches our numerical results within 6\% accuracy.

\subsection{PBH merger rate}
\label{subsec: merger}

The orbit of a PBH binary is parametrized by $j$, $r_a$, and two PBH masses.
For PBHs with monochromatic mass $m$, the merger rate $R$ is given by 
\begin{align}
    R(t)
    =
    \int \mathrm{d}t' \mathrm{d}j \mathrm{d}r_a
    \frac{\mathrm{d}^2R_b(t')}{\mathrm{d}j\mathrm{d}r_a}
    \delta(t-t'-\tau(m,j,r_a)),
    \label{eq: merger rate original}
\end{align}
where $R_b(t)$ is the binary formation rate at the cosmic time $t$.

When we focus on a given PBH, the distribution of the nearest PBH with respect to the separation $x_\mathrm{in}$, under the condition that the next nearest PBH lies farther than $y$, is obtained as
\begin{align}
    \mathrm{d}n_2(x_\mathrm{in},y)
    &=
    n_\mathrm{PBH}(x_\mathrm{in}) e^{-\Gamma(y)}
    \mathrm{d}V(x_\mathrm{in})
    ,
\end{align}
where
\begin{align}
    \Gamma(y)
    &\equiv
    \int_0^{V(y)} \mathrm{d}V(x) \,
    n_\mathrm{PBH}(x),
\end{align}
and $n_\mathrm{PBH}(x_\mathrm{in})$ is the comoving number density of the PBHs including their spatial correlations.
Here, we introduced $y$ as a parameter to avoid the three-body system in the following discussion.
Then, we can write the number-density distribution of the PBH pair with respect to the initial comoving separation $x_\mathrm{in}$ as
\begin{align}
    \mathrm{d}n_\mathrm{pair}(x_\mathrm{in},y)
    =
    \frac{1}{2} \bar{n} \mathrm{d}n_{2}(x_\mathrm{in},y),
\end{align}
where $\bar{n}$ is the average PBH number density, and the factor of $1/2$ is included to avoid double counting of binaries.
This probability provides the binary formation rate as
\begin{align}
    \label{eq:prob_dist_binary_formation}
    \frac{\mathrm{d}^2R_b(t)}{\mathrm{d}j\mathrm{d}r_a}
    =
    \int \mathrm{d}x_\mathrm{in}
    \frac{\mathrm{d}n_\mathrm{pair}(x_\mathrm{in},y)}{\mathrm{d}x_\mathrm{in}}
    \frac{\mathrm{d}^2P(j,r_a|x_\mathrm{in},y)}{\mathrm{d}j \mathrm{d}r_a}
    \delta(t-t_\mathrm{dc}(x_\mathrm{in})),
\end{align}
where $\mathrm{d}^2P (j, r_a|x_\mathrm{in}, y)/\mathrm{d}j \mathrm{d}r_a$ is the conditional probability distribution for the angular momentum $j$ and the semi-major axis $r_a$ of a PBH binary, given that it is formed at the initial separation $x_\mathrm{in}$ without extra PBHs inside the distance $y$ from the binary.
Since we are interested in $\tau \simeq t_0\gg t_\mathrm{eq}\gtrsim t_\mathrm{dc}(x_\mathrm{in})$, the merger rate at present is estimated as
\begin{align}
    R^{(y)}(t_0)
    &=
    \int \mathrm{d}t \mathrm{d}j \mathrm{d}r_a 
    \frac{\mathrm{d}^2R_b(t)}{\mathrm{d}j\mathrm{d}r_a}
    \delta(t_0-t-\tau(m_1,m_2,r_a,j))
    \nonumber \\
    &\simeq
    \int \mathrm{d} n_\mathrm{pair}(x_\mathrm{in},y) \mathrm{d}j
    \frac{\mathrm{d}P(j,r_a|x_\mathrm{in},y)}{\mathrm{d}j}
    \delta(t_0-\tau(m_1,m_2,r_a,j))
    \label{eq: merger rate mid}
    \\
    &=
    \frac{1}{7t_0}\int \mathrm{d}n_\mathrm{pair}(x_\mathrm{in},y)\,
    j \left.\frac{\mathrm{d}P(j,r_a|x_\mathrm{in},y)}{\mathrm{d}j}
    \right|_{j=j(r_a(x_\mathrm{in}),t_0)},
    \label{eq: merger rate}
\end{align}
where we used the fact that $r_a$ is given as a function of $x_\mathrm{in}$ (see Eq.~\eqref{eq: r_a}) in the second equality and the relation~\eqref{eq: coalescence} in the last equality.

In addition to Eq.~\eqref{eq: merger rate}, we have to take into account two effects that can destroy binaries.
One is the binary disruption caused by accidental encounters with other PBHs and gravothermal collapse of PBH halos, which we will discuss later.
The other is the binary disruption by the nearby PBH, which is taken into account by an appropriate choice of $y$.
To determine $y$, let us consider the PBH three-body system formed after the PBH binary formation.
Considering the decoupling from the Hubble flow, we expect that the three-body system is formed at $a=a_\mathrm{eq}/\delta_\mathrm{NN}$, where the $\delta_\mathrm{NN}$ is the overdensity due to a binary in the volume within distance $x_\mathrm{NN}$ from the binary and is given by
\begin{align}
    \delta_\mathrm{NN}
    \simeq 
    \frac{M-\bar{\rho}_\mathrm{PBH,0}V(x_\mathrm{NN})}{\bar{\rho}_{\mathrm{m},0}V(x_\mathrm{NN})}.
\end{align}
From the above equation, we obtain $\bar{N}(x_\mathrm{NN}) \equiv \bar{n}V(x_\mathrm{NN})$ as
\begin{align}
    \bar{N}(x_\mathrm{NN})
    =
    \frac{M}{m}\frac{f_\mathrm{PBH}}{f_\mathrm{PBH}+\left(\Omega_\mathrm{m}/\Omega_\mathrm{DM}\right)\delta_\mathrm{NN}} .
\end{align}
The three-body system can be formed during the radiation-dominated era if $\delta_\mathrm{NN}\gtrsim \delta_\mathrm{th}$, which leads to the upper bound on $\bar{N}(x_\mathrm{NN})$ and hence $x_\mathrm{NN}$.
To avoid the formation of a three-body system, we require that the nearby PBH should not exist within the upper bound of $x_\mathrm{NN}$.
Thus, to exclude all three-body channel mergers, we choose $y$ as
\begin{align}
    \bar{N}(y)
    =
    \frac{M}{m}\frac{f_\mathrm{PBH}}{f_\mathrm{PBH}+\left(\Omega_\mathrm{m}/\Omega_\mathrm{DM}\right)\delta_\mathrm{th}}
    .
\end{align}

\section{Angular momentum distribution of PBH binaries}
\label{sec: est}

As seen in Eq.~\eqref{eq: merger rate}, the distribution of binary angular momentum is a key ingredient for evaluating the PBH merger rate.
The angular momentum of a PBH binary is gravitationally induced by the surrounding PBHs and density perturbations of the matter component other than PBHs. 
Here, we divide the angular momentum $\bm{j}$ of a binary into these two contributions: one from isocurvature perturbations by surrounding PBHs, denoted as $\bm{j}_\mathrm{iso}$, and the other from adiabatic perturbations of matter, denoted as $\bm{j}_\mathrm{ad}$. 

Here, we compute the generating function of the angular momentum,
\begin{align}
\begin{aligned}
    Z(\bm{k})
    &\equiv 
    \int \mathrm{d}^3\bm{j}~
    e^{i\bm{k}\cdot\bm{j}}~\frac{\mathrm{d}^3P}{\mathrm{d}j^3}(\bm{j})\\
    &=\left<e^{i\bm{k}\cdot\bm{j}}\right>
    =\left<e^{i\bm{k}\cdot(\bm{j}_\mathrm{ad}+\bm{j}_\mathrm{iso})}\right>
    =e^{K_\mathrm{ad}(\bm{k})+K_\mathrm{iso}(\bm{k})}
    ,
\end{aligned}
\end{align}
where $\langle \ldots \rangle$ takes the average over all the random quantities inside.
The distribution of binary angular momentum is written as
\begin{align}
    \frac{\mathrm{d}^3P}{\mathrm{d}j^3}(\bm{j})
    =
    \int \frac{\mathrm{d}^3\bm{k}}{(2\pi)^3}~
    e^{-i\bm{k}\cdot\bm{j}}~Z(\bm{k}).
\end{align}
From the definition of $\bm{j}$, physical angular momenta satisfy $\bm{r}\cdot\bm{j}=0$.
To exploit this property, we decompose $\bm{k}$ as $\bm{k} =\bm{k}_\perp + \bm{k}_\parallel$ satisfying $\bm{k}_\perp\cdot\bm{r}=0$ and $\bm{k}_\parallel\times\bm{r}=0$.
Similarly, we also decompose $\bm{j} = \bm{j}_\perp + \bm{j}_\parallel$, where $\bm{j}_\parallel$ should vanish for physical angular momenta.
Then, we obtain $\bm{k} \cdot \bm{j} = \bm{k}_\perp \cdot \bm{j}_\perp$, implying that $K$ is given as a function of $\bm{k}_\perp$, i.e., $K(\bm{k}) = K(\bm{k}_\perp)$.
Consequently, we can rewrite the probability distribution as
\begin{align}
    \frac{\mathrm{d}^3P}{\mathrm{d}j^3}
    =
    \int\frac{\mathrm{d}^3\bm{k}}{(2\pi)^3}
    e^{-i\bm{k}_\perp\cdot\bm{j}+K(\bm{k}_\perp)}
    e^{-i\bm{k}_\parallel\cdot\bm{j}}
    =
    \delta(\bm{j}_\parallel) \int\frac{\mathrm{d}^2\bm{k}_\perp}{(2\pi)^2}
    e^{-i\bm{k}_\perp\cdot\bm{j}+K(\bm{k}_\perp)}.
    \label{eq: cumulant}
\end{align}
Since the system is statistically symmetric under rotations around the $\bm{r}$ direction, we can denote $K(\bm{k})$ by $K(|\bm{k}|)$.

\subsection{Contribution from matter density fluctuation}
\label{subsec: matter}
Here, we first calculate $\left<|\bm{j}_\mathrm{ad}|^2\right>$.
The PBH binary located at $\bm{r}_c$ is subject to tidal forces $T_{ij}$ from the surrounding density perturbations, which is given by
\begin{align}
    T_{ij}(\bm{r}_c)
    =
    \partial_i\partial_j
    \left(
        - G \int \mathrm{d}^3\bm{r'}
        \frac{\rho_\mathrm{m}(\bm{r}')}{|\bm{r}_c-\bm{r}'|}
    \right). 
    \label{eq: torque}
\end{align}
Using the comoving momentum $\bm{q}$, the Fourier transformation of Eq.~\eqref{eq: torque} is written as
\begin{align}
    T_{ij}(\bm{q})
    =
    4\pi G a^{-3} 
    \rho_\mathrm{m}(\bm{q})
    \hat{\bm{q}}_i\hat{\bm{q}}_j,
\end{align}
where $\hat{\bm{q}}\equiv \bm{q}/|\bm{q}|$, and $\rho_\mathrm{m}(\bm{q})$ is the Fourier transform of $\rho_\mathrm{m}(\bm{r})$ defined by $\rho_\mathrm{m}(\bm{q})\equiv\int \mathrm{d}^3\bm{r}' \rho_\mathrm{m}(\bm{r}')\\\exp{\left(-i\frac{\bm{q}}{a}\cdot\bm{r}'\right)}$.
The angular momentum induced by the matter density fluctuations is
\begin{align}
    |\bm{j}_\mathrm{ad}(\bm{q})|^2
    =
    \left(0.95\frac{x_\mathrm{in}^3}{GM}\right)^2\left(4\pi G\rho_\mathrm{m}(\bm{q})\right)^2
    |\hat{\bm{x}}\times\hat{\bm{q}}|^2
    (\hat{\bm{q}}\cdot\hat{\bm{x}})^2 
    ,
\end{align}
where $\hat{\bm{x}}=\bm{x}/\lvert\bm{x}\rvert$.
By averaging over $\hat{\bm{x}}$, we obtain
\begin{align}
    \left<|\bm{j}_\mathrm{ad}(\bm{q})|^2\right>
    =
    \frac{6}{5}j_\mathrm{ch}^2\left(\frac{\Omega_{\mathrm{m}}}{\Omega_\mathrm{PBH}}\right)^2 \left<\delta_\mathrm{ad}^2(\bm{q})\right> 
    \equiv 
    \frac{6}{5} j_\mathrm{ch}^2 \frac{\sigma_\mathrm{m}^2}{f_\mathrm{PBH}^2},
\end{align}
where $\delta_\mathrm{ad}(\bm{q}) \equiv \rho_\mathrm{m}(\bm{q})/\bar{\rho}_\mathrm{m}$ 
is the density contrast of the momentum scale $q$, and $j_\mathrm{ch}$ is the characteristic angular momentum defined by
\begin{align}
    j_\mathrm{ch} \equiv 0.95 \bar{N}(x_\mathrm{in})\frac{m}{M}.
    \label{eq : def of j0}
\end{align}
Here we set the value of $\left<\delta_\mathrm{ad}^2(\bm{q})\right>$ around the PBH scale, which is given by the extrapolation from the CMB observation \cite{Ali-Haimoud:2017rtz}
\begin{align}
    \left<\delta_\mathrm{ad}^2(\bm{q})\right>
    =0.005^2 .
\end{align}
Next, we discuss the probability distribution of $\bm{j}_\mathrm{ad}$.
Assuming the Gaussian density fluctuations, the distribution of the angular momentum induced by matter fluctuations is given by
\begin{align}
    P(\bm{j}_\mathrm{ad})
    =
    \frac{1}{\pi \sigma_{j,\mathrm{ad}}^2}
    \exp{\left(
        -\frac{|\bm{j}_\mathrm{ad}|^2}
        {\sigma_{j,\mathrm{ad}}^2}
    \right)} , 
\end{align}
where
\begin{align}
    \sigma_{j,\mathrm{ad}}^2
    =
    \left< |\bm{j}_\mathrm{ad}|^2 \right> .
\end{align}
The cumulant is calculated as
\begin{align}
    K_\mathrm{ad}(\bm{k})
    =
    -\frac{1}{4}\sigma_{j,\mathrm{ad}}^2 
    |\bm{k}_\perp|^2
    . 
\end{align}

\subsection{Contribution from PBH density fluctuation}
\label{subsec: PBH}

Next, we discuss the probability distribution of $\bm{j}_\mathrm{iso}$.
The contributions to the angular momentum from the surrounding PBHs are written as
\begin{align}
    \bm{j}_\mathrm{iso}
    =
    \sum_k \bm{j}_1(\bm{x}_k,m_k) ,
\end{align}
where $\bm{j}_1$ is the contribution from a single PBH, $k$ is the label for PBHs surrounding the binary, and $\bm{x}_k$ and $m_k$ are the position and mass of the $k$-th PBH, respectively.
By definition,
\begin{align}
    T_{ij}(\bm{r}_c)
    =
    -\sum_k\partial_i\partial_j\frac{Gm_k}{|\bm{r}_c-\bm{r}_k|}
    =
    \sum_k\frac{Gm_k}{\left|\bm{r}_c-\bm{r}_k\right|^3}\left(\delta_{ij}-3\frac{(\bm{r}_c-\bm{r}_k)_i(\bm{r}_c-\bm{r}_k)_j}{\left|\bm{r}_c-\bm{r}_k\right|^2}\right).
\end{align}
When we take the coordinate with the PBH binary at the origin ($\bm{r}_c=0$), $T_{ij}$ is simplified as
\begin{align}
    T_{ij}
    =
    \sum_k\frac{Gm_k}{a^3\left|\bm{x}_k\right|^3}\left(\delta_{ij}-3(\hat{\bm{x}}_k)_i(\hat{\bm{x}}_k)_j\right) .
    \label{eq: from PBH}
\end{align}
Thus, we have the dimensionless angular momentum
\begin{align}
    \bm{j}_{1}(\bm{x}_k,m_k)
    &=
    -j_\mathrm{ch}\frac{m_k}{\left<m\right>}\frac{3}{\bar{N}(\bm{x}_k)}(\hat{\bm{x}}_k\times\hat{\bm{r}})(\hat{\bm{x}}_k\cdot\hat{\bm{r}}) 
    \nonumber \\
    &=
    -j_\mathrm{ch} \frac{3}{\bar{N}(\bm{x}_k)}(\hat{\bm{x}}_k\times\hat{\bm{r}})(\hat{\bm{x}}_k\cdot\hat{\bm{r}}) .
\end{align}
Here and hereafter, we assume monochromatic mass for PBHs and set the PBH mass to $30M_\odot$ when performing numerical evaluations.

The cumulant for the binary angular momentum is 
\begin{align}
    K_\mathrm{iso}(\bm{k})
    =
    N\ln{\left<e^{i\bm{k}\cdot\bm{j}_1}\right>}
    =
    N\ln{\left(\frac{1}{N}\int \mathrm{d}^3\bm{x}\,
    n_\mathrm{PBH}(\bm{x})e^{i\bm{k}\cdot\bm{j}_1(\bm{x})}\right)},
\end{align}
where $N \equiv \int \mathrm{d}^3\bm{x}\, n_\mathrm{PBH}(\bm{x})$ is the number of PBHs around the binary.
For a large $N$ limit with a fixed number density $\bar{n}$,
\begin{align}
    K_\mathrm{iso}(\bm{k})
    =
    \lim_{N\rightarrow\infty}
    N\ln{\left(1+\frac{K_0(\bm{k})}{N}\right)}
    =
    K_0(\bm{k}) ,
\end{align}
where
\begin{align}
    K_0(\bm{k})
    \equiv 
    \int \mathrm{d}^3\bm{x}\,
    n_\mathrm{PBH}(\bm{x})\left(e^{i\bm{k}\cdot\bm{j}_1(\bm{x})}-1\right).
\end{align}
Using $u\equiv z V(y)/V(x)$ and $z\equiv k_\perp j_\mathrm{ch}/\bar{N}(y)$, we obtain
\begin{align}
    K_0(\bm{k})
    =
    \bar{N}(y) z \int \frac{\mathrm{d}u}{u^2} 
    \frac{n_\mathrm{PBH}(x(u))}{\bar{n}} 
    \left[
        \frac{\pi}{2\sqrt{2}} J_{-\frac{1}{4}}\left(\frac{3}{4}u\right)
        J_{\frac{1}{4}}\left(\frac{3}{4}u\right) - 1
    \right]
    ,
\end{align}
where $n_\mathrm{PBH}(x) \equiv n_\mathrm{PBH}(\bm{x})$, and $J_\nu (x)$ is the Bessel function.

\subsection{Binary angular momentum for clustered PBHs}
\label{subsec: binary}
To describe the PBH clustering, we introduce the correlation function of PBHs, $\xi$, which is defined via the spatial distribution of PBHs around a single PBH located at the origin, $\bm{x} = \bm{0}$, as
\begin{align}
    n_\mathrm{PBH}(\bm{x})
    =
    \bar{n}(1+\xi(|\bm{x}|)).
    \label{eq: spatial}
\end{align}
For the sake of concreteness in the argument, we assume the power-law correlation function as
\begin{align}
    \xi(x)
    =
    \xi_c \left(\frac{x}{x_\ast}\right)^{-\alpha},
    \label{eq : functional form of xi}
\end{align}
where we set $x_\ast=1~\mathrm{Mpc}$.
This functional form provides a good approximation in concrete models of PBH formation from non-Gaussian density perturbations, in which $\alpha\sim1.2\,\text{--}\,1.5$ are predicted~\cite{Kawasaki:2021zir}.

We obtain the angular momentum distribution with respect to $j \equiv |\bm{j}|$ by the angular integral of Eq.~\eqref{eq: cumulant} as
\begin{align}
    \frac{\mathrm{d}P}{\mathrm{d}j}
    =
    \int \mathrm{d}\Omega_j \,
    j^2\frac{\mathrm{d}^3P}{\mathrm{d}j^3} ,
\end{align}
which can be rewritten as 
\begin{align}
    j\frac{\mathrm{d}P}{\mathrm{d}j}
    =
    j^2\int \mathrm{d}k_\perp \,
    k_\perp e^{K(k_\perp)} J_0(k_\perp j) .
\end{align}
Using $K_\mathrm{ad}(\bm{k})$ and $K_\mathrm{iso}(\bm{k})$ calculated in Secs.~\ref{subsec: PBH} and \ref{subsec: binary}, we obtain
\begin{align}
\begin{aligned}
    j\frac{\mathrm{d}P}{\mathrm{d}j}
    =&
    \int \mathrm{d}v\, v J_0(v) 
    \exp \Bigg[
        -v^2 \frac{3}{10} \frac{\sigma_\mathrm{m}^2}{f_\mathrm{PBH}^2} \frac{j_\mathrm{ch}^2}{j^2}
        \\
        &+v\frac{j_\mathrm{ch}}{j}
        \int\frac{\mathrm{d}u}{u^2}
        \left( 1 + \xi(x(u,v)) \right)
        \left( 
            \frac{\pi}{2\sqrt{2}}
            J_{-\frac{1}{4}} \left(\frac{3}{4}u \right)
            J_{\frac{1}{4}} \left(\frac{3}{4}u\right) - 1
        \right)
    \Bigg] ,
    \label{eq: angular distribution}
\end{aligned}
\end{align}
where $v \equiv k_\perp j$, and
\begin{align}
    x(u,v)
    \equiv 
    \left(\frac{3v}{4\pi u \bar{n}}\frac{j_\mathrm{ch}}{j}\right)^\frac{1}{3}.
\end{align}

Note that Eq.~\eqref{eq: angular distribution} is a function of $j/j_\mathrm{ch}$ and has a peak at $j/j_\mathrm{ch}\sim \mathcal{O}(1)$ at least in the absence of clustering.%
\footnote{%
The typical value of $j/j_\mathrm{ch}$ could be larger for clustered PBHs.
Thus, the assumption of $j \sim j_\mathrm{ch}$ corresponds to an underestimate of the typical value of $j$, leading to an overestimate of the effects of binary disruption, which we will discuss in Sec.~\ref{sec: suppression}. 
However, the binary disruption hardly affects the merger rate in the parameter range relevant to the merger rate inferred from the observations for the case with clustering, even if we assume $j \sim j_\mathrm{ch}$, as we will see below.
Thus, we adopt this assumption in both cases, with and without clustering.
}
By combining Eqs.~\eqref{eq: r_a},~\eqref{eq: coalescence}, and~\eqref{eq : def of j0}, $j_\mathrm{ch}$ can be expressed as a function of $j$ when the coalescence time $\tau$ is fixed. Solving $j/j_\mathrm{ch}=1$ with respect to $j$ under $\tau = t_0$ leads to $j= j_\tau$, where $j_\tau$ is given by 
\begin{align}
    j_\tau
    \equiv 
    1.9\times10^{-2} \left(\frac{M}{M_\odot}\right)^\frac{5}{37}\left(4 \frac{\mu}{M}\right)^\frac{3}{37} f_\mathrm{PBH}^\frac{16}{37}
    \ .
    \label{eq : def of jtau}
\end{align}
Here, $j_\tau$ can be understood as the typical angular momentum with a given coalescence time $\tau = t_0$.

\section{Disruption of binaries in the halos}
\label{sec: suppression}

Besides the formation of the three-body system due to the nearby PBH, there are two astrophysical contributions to the binary disruption.
One is the accidental encounter of PBHs inside a halo, and the other is the core collapse of a halo.
The PBH clustering affects both processes by changing the halo formation epochs and halo densities.
We first discuss the effects of clustering on the PBH encounter and the core collapse in turn.
Then, we discuss the halo formation and derive the suppression factor for the merger rate in the presence of PBH clustering.

\subsection{Binary disruption by astrophysical effects}
\label{subsec: astro}

We first discuss the accidental encounter of an external PBH.
Here, we start from the case in which a PBH with mass $m$ encounters a PBH binary with a total mass $M$.
Suppose that, in the binary rest frame, the PBH with the initial velocity $v_\mathrm{rel}$ and the impact parameter $b$ passes near the binary, and the distance to the binary and the velocity at the closest approach are $r_p$ and $v_p$, respectively.
From the conservation of angular momentum, we obtain
\begin{align}
    v_p=\frac{b}{r_p} v_\mathrm{rel} .
    \label{eq: momentum cons}
\end{align}
The scattering takes the time $t_p\sim r_p/v_p$.
During the scattering, Eq.~\eqref{eq: from PBH} induces the angular momentum of the PBH binary,
\begin{align}
    \Delta\bm{L}
    \simeq
    - \mu t_p \bm{r}_a\times T_p\bm{r}_a ,
\end{align}
where
\begin{align}
    (T_p)_{mn}
    =
    \frac{G m}{r_p^3}(\delta_{mn}-3(\hat{\bm{r}}_p)_m(\hat{\bm{r}}_p)_n) .
\end{align}
The dimensionless angular momentum induced by this scattering process is estimated as
\begin{align}
    \Delta\bm{j}
    =
    \frac{G^\frac{1}{2} m r_a^\frac{3}{2}}{M^\frac{1}{2} r_p b v_\mathrm{rel}} .
\end{align}
The binary may be disrupted if $\Delta j\gtrsim j_\tau$ with $j_\tau$ defined in Eq.~\eqref{eq : def of jtau}.
Due to the large initial separation between the binary and external PBH, we have $v_\mathrm{rel}\ll v_p$.
The conservation of energy provides
\begin{align}
    r_p
    \simeq
    \frac{b^2v_\mathrm{rel}^2}{2 G (M+m)} .
\end{align}
Consequently, the binary disruption condition becomes
\begin{align}
    b
    \lesssim
    \frac{G^\frac{1}{2}m^\frac{1}{3}(M+m)^\frac{1}{3}r_a^\frac{1}{2}}{j_\tau^\frac{1}{3} M^\frac{1}{6} v_\mathrm{rel}}
    \equiv
    b_{j_\tau} ,
\end{align}
and the cross section for the binary disruption is $\sigma_{j_\tau}=\pi b_{j_\tau}^2$. 
Here, we calculate the binary lifetime in the halo composed of $N$ PBHs.
The spherical collapse model provides the halo density
\begin{align}
    \rho_H
    =
    18 \pi^2 \bar{\rho}_{\mathrm{m},0} a_H^{-3} ,
\end{align}
where $a_H$ is the scale factor at the halo formation time determined by the decoupling from the Hubble flow
\begin{align}
    a_H\sim\frac{a_\mathrm{eq}}{\delta_H},
\end{align}
and $\delta_H$ is the density contrast that forms the halo.
Based on the Gaussian approximation of the density perturbations, the typical value of $\delta_H$ is derived from the discussion of the power spectrum in the next subsection.
For given $\alpha$, using Eq.~\eqref{eq: PS form},
\begin{align}
    \delta_H
    \sim
    \sigma_R
    \sim
    \frac{\Omega_\mathrm{DM}}{\Omega_\mathrm{m}}f_\mathrm{PBH}\sqrt{c_\xi\xi_c+\frac{1}{N}} ,
\end{align}
where
\begin{align}
    c_\xi
    =
    \frac{2^{3-\frac{\alpha}{3}}3^{2-\frac{\alpha}{3}}\pi^\frac{\alpha}{3}}{-\alpha^3+13\alpha^2-54\alpha+72}
    \left(\frac{m N}{f_\mathrm{PBH} \rho_\mathrm{DM,0} x_\ast^{3}}\right)^{-\frac{\alpha}{3}}.
\end{align}
Here, we have replaced the density contrast by its dispersion $\sigma_R$.
We expect that the mass of the virialized halo with $N$ PBHs is given by
\begin{align}
    M_N 
    \equiv
    N m \frac{\Omega_\mathrm{m}}{f_\mathrm{PBH}\Omega_\mathrm{DM}}
    \ ,
\end{align}
and using this mass, the virial radius $R_\mathrm{vir}=(3M_N/4\pi\rho_H)^{1/3}$ provides the velocity dispersion $v_\mathrm{rel}\sim\sigma_v=\sqrt{GM_N/R_\mathrm{vir}}$.
Finally, we obtain the lifetime of a binary in a halo with $N$ PBHs as
\begin{align}
    t_\mathrm{bin}
    &=
    \frac{1}{(\Omega_\mathrm{DM}f_\mathrm{PBH}\rho_H/\Omega_\mathrm{m}m)\cdot v_\mathrm{rel}\cdot\sigma_{{j_\tau}}}
    \nonumber \\
    &\sim
    14\,\mathrm{Gyr}
    \bigg(\frac{1}{1+Nc_\xi\xi_c}\bigg)^\frac{5}{4}
    \bigg(\frac{N}{1000}\bigg)^\frac{19}{12} f_\mathrm{PBH}^{-\frac{619}{222}}\bigg(\frac{m}{M_\odot}\bigg)^{-\frac{10}{111}}
    .
\end{align}
Here, we have replaced $r_a$ by its typical value without clustering, which is obtained by substituting $j=j_\tau$ with $\tau=t_0$ into Eq.~\eqref{eq: coalescence}.
Here, we require $t_\mathrm{bin}$ to be larger than the present time for the binary to contribute to the present merger rate.
From this condition, we obtain the lower bound on $N$, which we denote by $N_\mathrm{enc}$.

Next, we consider the core collapse of the halo.
If there is a PBH binary in the halo, it can experience disruption through the core collapse.
According to Ref.~\cite{Quinlan:1996bw}, the typical timescale of the core collapse satisfies $t_\mathrm{cc}\geq18 t_r$, where the relaxation time $t_r$ is given by~\cite{Spitzer:1987degc}
\begin{align}
    t_r=0.065 \frac{\sigma_v^3}{G^2m \rho_H \ln{\Lambda}},
\end{align}
and it can be approximated with $\ln{\Lambda}\sim\ln \left( N/f_\mathrm{PBH}\right)$.
As the condition for the binary to be disrupted, we adopt the criteria $t_0\geq 18 t_r$ following Ref.~\cite{Raidal:2024bmm}.
In other words, we assume that the core collapse time is as short as $t_\mathrm{cc} \sim 18t_r$.
If the core collapse takes a longer time, the suppression of the merger rate due to the core collapse will be milder.
Using the virial theorem and the spherical collapse model, we obtain
\begin{align}
    t_r
    \sim
    5.8\,\mathrm{kyr} \bigg(\frac{1}{1+Nc_\xi\xi_c}\bigg)^\frac{3}{4}\frac{N^\frac{7}{4}}{f_\mathrm{PBH}^\frac{5}{2}\ln{\Lambda}},
    \label{eq: tr}
\end{align}
and thus binaries are disrupted if 
\begin{align}
    N\leq 840~(1+N c_\xi\xi_c)^\frac{3}{7}\ f_\mathrm{PBH}^\frac{10}{7}(\ln{\Lambda})^\frac{4}{7},
    \label{eq: Ncc condition}
\end{align}
from which the upper bound on $N$ denoted as $N_\mathrm{cc}$ is obtained.%
\footnote{%
    The numerical coefficients in Eqs.~\eqref{eq: tr} and \eqref{eq: Ncc condition} are different from those in Ref.~\cite{Raidal:2024bmm}. 
    This is partly because we assume that the halos contain baryons, PBHs, and other dark matter components in the same fraction as in the background, while they assume that the halos consist of PBHs and other dark matter components.
}

\begin{figure}[t]
    \centering
    \includegraphics[width=.75\textwidth ]{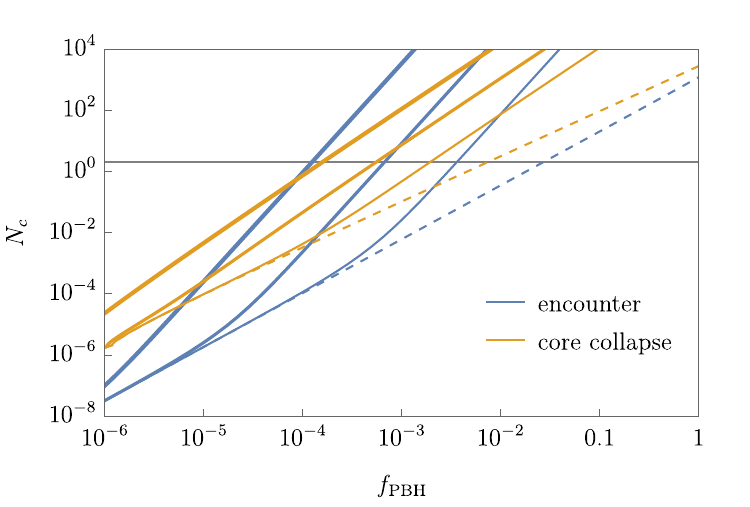}
    \caption{
        Estimate of $N_c$ as a function of $f_\mathrm{PBH}$ for $\alpha=1.5$.
        The blue and orange lines denote the estimates due to binary disruption by accidental encounters and halo core collapse, respectively.
        From thickest to thinnest, the solid lines correspond to $\xi_c=10^2,10^0,10^{-2}$ respectively, while the dashed line corresponds to $\xi_c=0$.
    }
    \label{fig: N_c}
\end{figure}

In Fig.~\ref{fig: N_c}, we compare the upper limits on the PBH number of a halo as a binary disruption condition.
A PBH binary in a halo with $N$ PBHs was disrupted before today if
\begin{align}
    N\leq N_c\equiv
    \max{\left[N_\mathrm{enc},N_\mathrm{cc}\right]}
    .
\end{align}
While the accidental encounters can take place in a halo with three or more PBHs, the core collapse can disrupt the binaries even in a halo with two PBHs.
Thus, we exclude PBH binaries in halos containing $N$ PBHs with $2 \leq N \leq N_c$ if $N_\mathrm{cc} \geq 2$, and exclude those with $3 \leq N \leq N_c$ otherwise.
We define the lower bound for this exclusion by $N_l \,(= 2 \text{\,or\,}3)$.
In Fig.~\ref{fig: N_c}, $N_\mathrm{cc} \geq 2$ always holds in the region where $N_c \geq 2$, except for the case of the strongest clustering, $\xi_c = 10^2$.
Although $N_c$ represents the critical number of PBHs in a halo and is expected to be an integer by definition, we allow non-integer values of $N_c$ here.
The corresponding treatment of $N_c$ in the evaluation of the suppression factor is discussed below.
In the next subsection, we discuss the probability that such configurations are realized.

\subsection{Probability that a PBH belongs to \texorpdfstring{$N$}{}-PBH halo}
\label{subsec: supp_prob}

Due to the PBH clustering, it is not appropriate to assume the Poisson distribution when we evaluate the probability that a PBH belongs to an $N$-PBH halo.
To calculate the probability, we generalize the method used in Ref.~\cite{Hutsi:2019hlw} to the PBH spatial distribution with two-point correlations.
The power spectrum of the density of PBHs with a monochromatic mass is given by
\begin{align}
    P_\mathrm{PBH}(\bm{k}) 
    =
    \frac{1}{\bar{n}} + 4 \pi \int^\infty_0 \mathrm{d}r\,
    r^2 \frac{\sin{k r}}{k r} \xi(r).
\end{align}
This is translated to the power spectrum of the matter density fluctuations as
\begin{align}
    P_\mathrm{m}(\bm{k}) = \left(\frac{\Omega_\mathrm{PBH}}{\Omega_\mathrm{m}}\right)^2 P_\mathrm{PBH}(\bm{k}).
\end{align}
Since the growth of the matter density perturbation after the matter radiation equality is written as
\begin{align}
    \delta(z)
    =
    D_+(z) \delta_i,\quad 
    D_+(z) 
    =
    {}_2F_1 \left(\frac{1-\sqrt{21}}{4},\frac{1+\sqrt{21}}{4};1;-\frac{1+z_\mathrm{eq}}{1+z}\right) ,
\end{align}
the power spectrum at the redshift $z$ is given by
\begin{align}
    P_\mathrm{m}(z,\bm{k}) = D_+^2(z) \left(\frac{\Omega_\mathrm{PBH}}{\Omega_\mathrm{m}}\right)^2 P_\mathrm{PBH}(\bm{k}).
\end{align}
Here ${}_2F_1$ is the hypergeometric function.

Let us consider the overdensity $\delta_R$ coarse-grained over the comoving length scale $R$ given by
\begin{align}
    R = \left(\frac{3 M_\mathrm{halo}}{4 \pi \bar{\rho}_{\mathrm{m},0}}\right)^\frac{1}{3} 
    .
\end{align}
Here, we approximate that $\delta_R$ follows the Gaussian distribution as
\begin{align}
    p(\delta_R;R,z) = \frac{1}{\sqrt{2 \pi \sigma^2}} \exp{\left(-\frac{\delta_R^2}{2 \sigma_R^2(z)}\right)}.
\end{align}
where the dispersion is defined by
\begin{align}
    \sigma_R^2(z)
    =
    \int\frac{\mathrm{d}^3 \bm{k}}{(2\pi)^3} W^2(k R)P_\mathrm{m}(z,\bm{k}) .
    \label{eq: PS form}
\end{align}
Here, the window function is given as 
\begin{align}
    W(k R) = \frac{3}{(kR)^2} \left(\frac{\sin{k R}}{k R}-\cos{k R}\right) .
\end{align}
In the Press-Schechter formalism~\cite{Press:1973iz}, the volume fraction of the halo of mass larger than $M_\mathrm{halo}$ is
\begin{align}
    \Pi_{M_\mathrm{halo}}(\delta_c;z) 
    =
    2 \int_{\delta_c}^\infty \mathrm{d}\delta\, p(\delta_R;R,z) ,
\end{align}
where $\delta_c=1.69$ is the linearly extrapolated critical overdensity for halo collapse.
Using the volume fraction, the number density of the halos is
\begin{align}
    \frac{\mathrm{d}n}{\mathrm{d}M_\mathrm{halo}}
    =
    -\frac{\bar{\rho}_{\mathrm{m},0}}{M_\mathrm{halo}} 
    \frac{\mathrm{d} \Pi_{M_\mathrm{halo}}}{\mathrm{d}M_\mathrm{halo}}
    .
\end{align}
The probability that the target PBH binary is included in halos with mass from $M_\mathrm{min}$ to $M_\mathrm{max}$ is given by
\begin{align}
    p_\mathrm{PBH}(M_\mathrm{max},M_\mathrm{min};z)
    =
    \dfrac{-\int^{M_\mathrm{max}}_{M_\mathrm{min}}
    \mathrm{d} M\,\dfrac{\mathrm{d} \Pi_M}{\mathrm{d}M}}
    {-\int^\infty_{M_2} \mathrm{d} M\,\dfrac{\mathrm{d} \Pi_{M}}{\mathrm{d}M}}
    =
    \frac{\Pi_{M_\mathrm{min}} - \Pi_{M_\mathrm{max}}}{\Pi_{M_2}}
    .
    \label{eq : prob in halo}
\end{align}
Also, the probability that the PBH binary is included in a subhalo with mass from $M_\mathrm{min}^\mathrm{sub}$ to $M_\mathrm{max}^\mathrm{sub}$ inside a halo with mass at least $M_\mathrm{min}$ is given by
\begin{align}
    p_\mathrm{PBH}^\mathrm{sub}(M^\mathrm{sub}_\mathrm{max},M^\mathrm{sub}_\mathrm{min};M_\mathrm{min};z)
    &=
    -
    \frac{1}{\Pi_{M_2}}
    \int_{M_\mathrm{min}}^\infty \mathrm{d} M_\mathrm{halo}\,\frac{\mathrm{d} \Pi_{M_\mathrm{halo}}}{\mathrm{d}M_\mathrm{halo}}
    \frac{p_\mathrm{PBH}(M^\mathrm{sub}_\mathrm{max},M^\mathrm{sub}_\mathrm{min};z)}{p_\mathrm{PBH}(M_\mathrm{halo},M_2;z)} 
    \nonumber \\
    &=
    \frac{\Pi_{M^\mathrm{sub}_\mathrm{min}} - \Pi_{M^\mathrm{sub}_\mathrm{max}}}{\Pi_{M_2}}
    \log \frac{\Pi_{M_2}}{\Pi_{M_2} - \Pi_{M_\mathrm{min}}} 
    .
    \label{eq : prob in subhalo}
\end{align}

\begin{figure}[t]
    \centering
    \includegraphics[width=.75\textwidth ]{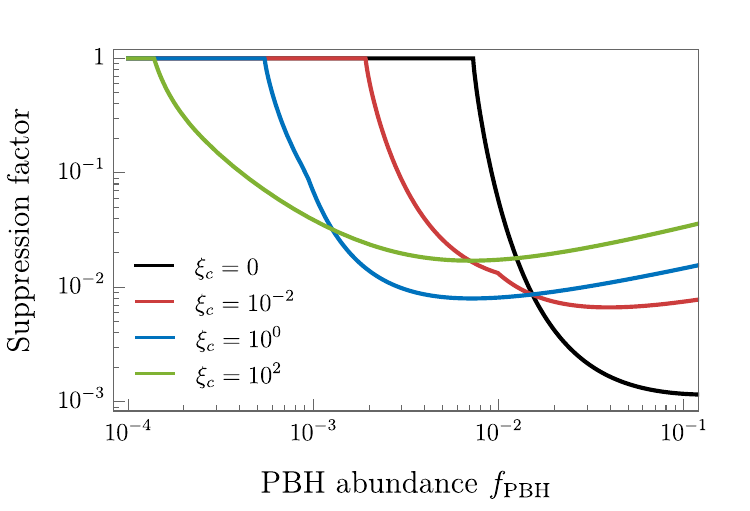}
    \caption{
        Suppression factor for the merger rate of PBH binaries.
    }
    \label{fig: SL}
\end{figure}

If a PBH binary belongs to a halo containing between $N_l$ and $N_c$ PBHs, it does not contribute to the present merger rate due to binary disruption.
In addition, even if a PBH binary belongs to a halo containing more than $N_c$ PBHs, it should be excluded from the merger-rate evaluation if it is contained in a subhalo with a PBH number between $N_l$ and $N_c$.
As a result, the suppression factor $S_L$ is given by~\cite{Raidal:2024bmm}
\begin{align}
    S_L
    &=
    1 - p_\mathrm{PBH}(M_{N_c},M_l;z_c) 
    - p^\mathrm{sub}_\mathrm{PBH}(M_{N_c},M_l;M_{N_c};z_c)
    \nonumber \\
    &=
    1 - \left(\frac{
    \Pi_{M_l} - \Pi_{M_{N_c}}}
    {\Pi_{M_{2}}}
    \right)
    \left[
    1 + \log \left(
    \frac{\Pi_{M_{2}}}
    {\Pi_{M_{2}} - \Pi_{M_{N_c}}}
    \right)\right]
    ,
    \label{eq : suppression factor}
\end{align}
for $N_c > N_l$.
Otherwise, binary disruption does not occur and $S_L = 1$.
Here, $z_c$ is the typical redshift at the formation of halos including $N_c$ PBHs.
Consequently, we obtain the suppression factor for the merger rate of PBHs, which is shown in Fig.~\ref{fig: SL}.
When $N_c<2$, the suppression factor equals unity by definition.
We can read the threshold value of $f_\mathrm{PBH}$ for $N_\mathrm{cc} \geq 2$ or $N_\mathrm{enc} \geq 3$ from Fig.~\ref{fig: N_c}.
In the case of $\xi_c = 0$, $N_\mathrm{cc}$ exceeds $2$ at $f_\mathrm{PBH} \simeq 10^{-2}$, above which $S_L$ deviates from unity in Fig.~\ref{fig: SL}, and the cases with clustering start to be suppressed below $f_\mathrm{PBH}=10^{-2}$ as expected from Fig.~\ref{fig: N_c}.
In addition, PBH clustering suppresses the formation of smaller halos that lead to binary disruption, and therefore we obtain larger $S_L$ for larger $\xi_c$ with $f_\mathrm{PBH} \gtrsim 10^{-2}$.
Due to the approximation of the Gaussian density perturbation, our suppression factor does not perfectly match the previous calculation using Poisson statistics for the case without clustering (see Appendix~\ref{app: consis check}). 
While $p_\mathrm{PBH}$ is in good agreement with the previous work~\cite{Raidal:2024bmm} in the limit of large halo with $f_\mathrm{PBH}\sim1$, our result of $S_L$ deviates from the previous work by a factor of at most $5$ for $f_\mathrm{PBH} = \mathcal{O}(10^{-2} \text{\,--\,} 10^{-1})$ (see Fig.~\ref{fig: result} below).
In addition, as seen later, the suppression factor is not effective for the merger rate inferred by the LVK observations in the cases with clustering.
Thus, our Gaussian approximation for the PBH distribution can be applied to clustering PBHs without causing significant errors.

Before closing this section, we comment on our treatment of non-integer values of $N_c$.
Since $N_c$ physically represents the number of PBHs in a halo, it might be natural to impose $N_c$ to be an integer, e.g., by replacing $N_c$ with its integer part. 
With such a prescription, $S_L$ becomes a discontinuous function of $f_\mathrm{PBH}$, reflecting the step-like change of $N_c$.
We regard this discontinuity as an artifact of the simplified assumptions adopted in our estimate, such as the one-to-one correspondence between halo mass and PBH number, and the use of a sharp threshold for binary disruption. 
In a more realistic situation, the PBH number should fluctuate in halos with the same mass, and binary disruption could occur even in halos containing PBHs slightly more than $N_c$.
As a result, $S_L$ is expected to depend smoothly on $f_\mathrm{PBH}$. 
We effectively realize the continuous dependence of $S_L$ on $f_\mathrm{PBH}$ by allowing non-integer values of $N_c$.

\section{Results of merger rate calculations}
\label{sec: res}

Here, we evaluate the merger rate derived in Secs.~\ref{sec: rev} and \ref{sec: est}:
\begin{align}
    R 
    &=
    S_L \frac{\bar{n}}{14 \tau} 
    \int \mathrm{d}x_\mathrm{in} \int \mathrm{d}u \,
    4\pi x_\mathrm{in}^2 
    e^{-\Gamma(y)} \left( 1 + \xi(x_\mathrm{in}) \right) 
    \bar{n} \, u \, J_0(u) 
    \nonumber \\
    &\quad \times \exp\Bigg[ 
      - \frac{3}{10} \frac{\sigma_\mathrm{m}^2}{f_{\text{PBH}}^2} \frac{j_\mathrm{ch}^2}{j^2} u^2 
    \nonumber \\
    &\quad\quad 
      - \bar{N}(y) \bigg( 
        -1 + {}_1F_2\left(
          -\frac{1}{2}; \frac{3}{4}, \frac{5}{4}; 
          -\frac{9}{16} \left( \frac{u j_\mathrm{ch}}{\bar{N}(y) j} \right)^2 
        \right)
        \nonumber \\
    &\quad\quad\quad 
        + \frac{\xi_c}{1 - \frac{\alpha}{3}} \left( 
            \frac{3m \bar{N}(y)}{4\pi \bar{\rho}_{\text{PBH}}x_\ast^3} 
          \right)^{-\frac{\alpha}{3}} 
          \cdot \bigg[
            -1 + {}_2F_3\left(
              \frac{1}{2}, -\frac{1}{2} + \frac{\alpha}{6};
              \frac{3}{4}, \frac{5}{4}, \frac{1}{2} + \frac{\alpha}{6};
              -\frac{9}{16} \left( \frac{u j_\mathrm{ch}}{\bar{N}(y) j} \right)^2 
            \right)
          \bigg]
      \bigg)
    \Bigg],
\end{align}
where ${}_1 F_2$ and ${}_2 F_3$ are generalized hypergeometric functions.
This is a generalized form of the merger rate obtained in Ref.~\cite{Raidal:2024bmm}.

\begin{figure}[t]
    \centering
    \includegraphics[width=.75\textwidth ]{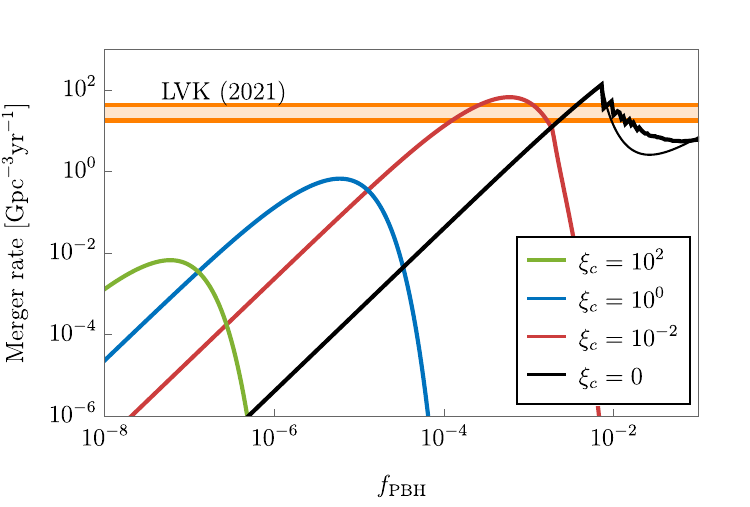}
    \caption{
        The merger rate of clustered PBHs for $m = 30M_\odot$ and $\alpha=1.5$.
        The solid line shows our calculation result, and the dashed line shows the previous work \cite{Kawasaki:2021zir}.
        The black, red, blue, and green lines denote the different strengths of the correlation $\xi_c=0,~10^{-2},~10^0,~10^2$, respectively.
        The orange shaded band represents the merger rate inferred from LVK gravitational-wave observations \cite{KAGRA:2021duu}.
    }
    \label{fig: result}
\end{figure}

In Fig.~\ref{fig: result}, we show the merger rate for PBHs with the monochromatic mass $m = 30M_\odot$ and different magnitudes of PBH clustering.
In the case without clustering (black), the merger rate is suppressed due to the astrophysical effects in halos for $f_\mathrm{PBH}\gtrsim10^{-2}$. 
Regarding the suppression factor for the no-clustering case, we also show the evaluation based on the Poisson statistics in Ref.~\cite{Raidal:2024bmm} with the thick black line, while our results are shown with the thin line.
On the other hand, in the presence of initial clustering, the merger rate is exponentially suppressed due to the formation of three-body systems even for $f_\mathrm{PBH}$ with which the astrophysical suppression is absent.
In other words, the suppression factor $S_L$ deviates from unity only when the suppression due to three-body system formation becomes significant in the cases with clustering.
We also show the inferred merger rate from the LVK observation with the orange band.
Consequently, we find that the inferred merger rate can be realized in the cases without clustering and with relatively weak initial clustering, $\xi_c = 10^{-2}$.
While the suppression due to astrophysical binary disruption plays an important role in determining the favored value of $f_\mathrm{PBH}$ in the absence of clustering, it does not affect the determination of $f_\mathrm{PBH}$ in the cases with clustering.
In particular, the precise evaluation of the suppression factor is essential in the case without clustering, which is beyond the scope of this work (see also Appendix~\ref{app: consis check}).
The qualitative behavior of the merger rates obtained here is consistent with those in previous work~\cite{Kawasaki:2021zir}, while the treatment of binary angular momenta is different from ours there.
See Appendix~\ref{app: comparison} for a detailed comparison with previous work~\cite{Kawasaki:2021zir}.

\section{Conclusions and Discussions}
\label{sec: disc}

Primordial black holes can form binaries in the early universe and merge at the present time, providing a possible probe of the early universe through gravitational-wave observations.
While the merger rate of PBH binaries was conventionally evaluated assuming a uniform initial PBH distribution, initial clustering can significantly affect the binary formation process and the subsequent merger rate.
In particular, clustering modifies the distribution of the initial PBH separation and the induced angular momentum of binaries, and can also influence the formation of PBH halos, which potentially affects astrophysical binary disruption.

In this work, we study the merger rate of PBH binaries in the presence of initial clustering by extending the existing formalism developed for Poisson-distributed PBHs.
The initial clustering of PBHs modifies the local PBH density.
In the merger rate evaluation, it changes the realization probability of the early two-body channel and the angular momentum contribution from surrounding PBHs.
In this work, we statistically estimate the binary angular momentum considering the contribution from all external PBHs and the matter density fluctuations, incorporating the effects of initial clustering.
The effects of initial clustering are also taken into account in evaluating the astrophysical binary disruption.
Moreover, we also improve the evaluation by refining the criterion for binary formation.
Using this formalism, we evaluated the present merger rate of clustered PBHs with the mass of $30 M_\odot$. 
Consequently, we found that, with relatively weak clustering, the LVK events can be explained with a smaller value of $f_{\rm PBH}$ than in the case without clustering in the early two-body channel.
On the other hand, with stronger clustering, the merger rate in the two-body channel is significantly suppressed due to the formation of three-body systems.
This suggests the possibility that the PBH mergers in three-body or many-body channels can play an important role in explaining the observed merger events, which is a promising direction of future studies.

Due to the several differences in the treatment of angular momentum and the formation conditions of binaries and three-body systems, the resulting merger rate differs from those in previous work~\cite{Raidal:2024bmm,Kawasaki:2021zir}.
Our result shows that the additional contribution to the angular momentum from outer PBHs and matter fluctuations influences the merger rate at small $f_\mathrm{PBH}$ tail, although it is almost irrelevant to the merger rate inferred from the LVK results (see also Appendix~\ref{app: comparison}).
We also modified the formation condition of three-body systems, which changes the exponential suppression at large $f_\mathrm{PBH}$ tail.
We can see this trend by comparing Figs.~\ref{fig: result} and \ref{fig: strong clust}.
We summarize the setups for the binary formation condition, the three-body formation condition, and the binary angular momentum in this work and previous work in Table~\ref{table: summary}. 

\begin{table}[t]
  \centering
  \caption{Summary of the setups in terms of the binary formation condition, the three-body system formation condition, and binary angular momentum.}
  \label{table: summary}
\begin{tikzpicture}[every path/.style={line width=0.4pt}]
  \def\cellwidth{3.5cm}
  \def\cellheight{1cm}

  \def\tablewidth{4*\cellwidth}
  \def\tableheight{4*\cellheight}

  \draw (0,0) rectangle (\tablewidth, -\tableheight);

  \draw[double, double distance=0.8pt] (0, -\cellheight) -- (\tablewidth, -\cellheight); 
  \draw (0, -2*\cellheight) -- (\tablewidth, -2*\cellheight); 
  \draw (0, -3*\cellheight) -- (\tablewidth, -3*\cellheight); 

  \draw[double, double distance=0.8pt] (\cellwidth, 0) -- (\cellwidth, -\tableheight); 
  \draw (2*\cellwidth, 0) -- (2*\cellwidth, -\tableheight); 
  \draw (3*\cellwidth, 0) -- (3*\cellwidth, -\tableheight); 

  \draw (0, 0) -- (\cellwidth, -\cellheight);

  \node at (1.5*\cellwidth, -0.5*\cellheight) {binary};
  \node at (2.5*\cellwidth, -0.5*\cellheight) {three body};
  \node at (3.5*\cellwidth, -0.5*\cellheight) {angular momentum};

  \node at (2.5*\cellwidth, -1.5*\cellheight) {$\delta_\mathrm{NN}>\sigma_m$};
  \node at (0.5*\cellwidth, -1.5*\cellheight) {Ref.~\cite{Raidal:2024bmm}};
  \node at (3.5*\cellwidth, -1.25*\cellheight) {all PBHs};
   \node at (3.5*\cellwidth, -1.75*\cellheight) {matter fluctuations};
  \node at (1.5*\cellwidth, -1.5*\cellheight) {$\delta_\mathrm{pair}>1-\frac{\Omega_\mathrm{PBH}}{\Omega_\mathrm{m}}^\ast$};

  \node at (2.5*\cellwidth, -2.5*\cellheight) {$\delta_\mathrm{NN}>1$};
  \node at (1.5*\cellwidth, -2.5*\cellheight) {$\delta_\mathrm{pair}>1$};
  \node at (0.5*\cellwidth, -2.5*\cellheight) {Ref.~\cite{Kawasaki:2021zir}};
  \node at (3.5*\cellwidth, -2.5*\cellheight) {nearest PBH};

  \node at (2.5*\cellwidth, -3.5*\cellheight) {$\delta_\mathrm{NN}>\delta_\mathrm{th}$};
  \node at (1.5*\cellwidth, -3.5*\cellheight) {$\delta_\mathrm{pair}>\delta_\mathrm{th}$};
  \node at (0.5*\cellwidth, -3.5*\cellheight) {this work};
  \node at (3.5*\cellwidth, -3.25*\cellheight) {all PBHs};
    \node at (3.5*\cellwidth, -3.75*\cellheight) {matter fluctuations};
\end{tikzpicture}
\\
\raggedright
\small{${}^\ast$Although the binary formation condition is discussed in Ref.~\cite{Raidal:2024bmm}, it is not taken into account in the merger rate calculation (see Appendix~\ref{app: wo clust consis check}).}
\end{table}

We found that astrophysical disruption of binaries can be relevant for merger rates in the range inferred by the LVK observations when the initial clustering is absent or relatively weak.
This highlights the importance of accurately evaluating the suppression factor in order to determine the PBH abundance for a given clustering strength.
In the present analysis, the suppression factor is estimated based on a rough treatment of the angular momentum induced by accidental encounters and of the timescale of the core-collapse process.
In addition, we assume a one-to-one correspondence between the halo mass and the number of PBHs in the halo.
Although refining these aspects will be important for further progress in this subject, such improvements would require dedicated numerical simulations of halo formation and are beyond the scope of this work.

\begin{acknowledgments}
This work was supported by JSPS KAKENHI Grant Numbers 23KJ0088 (K.M.), 24K17039 (K.M.), 25K07297 (M.K.) 25KJ1164(K.K.), and JST SPRING, and Grant Number JPMJSP2108(K.K. and S.N.).
This work is also supported by Grant-in-Aid for JSPS Research Fellow JP25KJ1030 (S.N.), and World Premier International Research Center
Initiative (WPI), MEXT, Japan (M.K.).
\end{acknowledgments}

\appendix

\section{Fitting formulae for binary formation}
\label{app: valid}

\begin{figure}[t]
    \centering
    \includegraphics[width=.7\textwidth ]{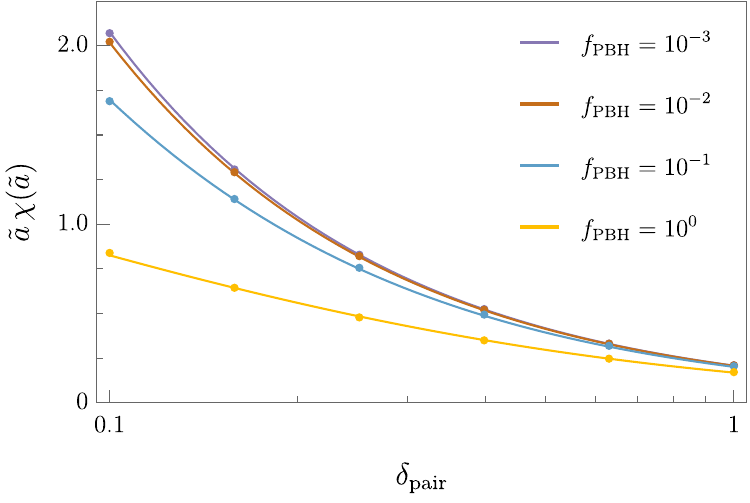}
    \caption{
        Amplitude of the radial separation of the binary.
        The purple, brown, light blue, and yellow dots denote the numerical results for $f_\mathrm{PBH}=10^{-3},10^{-2},10^{-1}$, and $10^0$, respectively.
        The lines show the fitting formula~\eqref{eq: binary_amp} with the corresponding value of $\delta_\mathrm{pair}$ and $f_\mathrm{PBH}$.
    }
    \label{fig: bin_form_amp}
\end{figure}
\begin{figure}[t]
    \centering
    \includegraphics[width=.7\textwidth ]{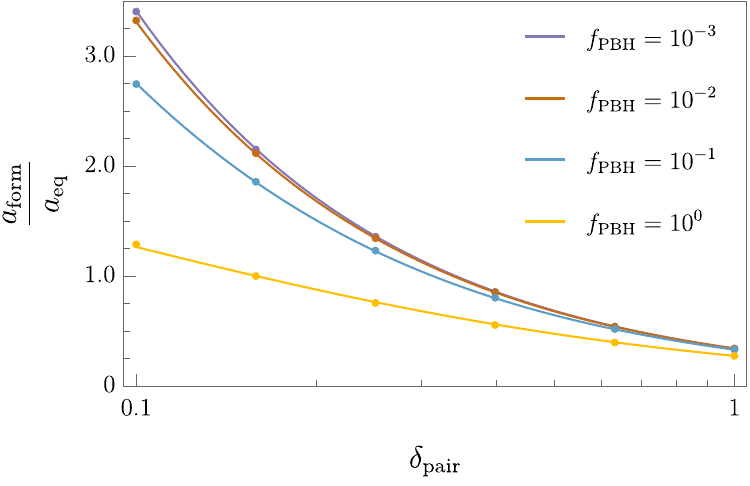}
    \caption{
        Scale factor at the binary formation.
        The purple, brown, light blue, and yellow dots denote the numerical results for $f_\mathrm{PBH}=10^{-3},10^{-2},10^{-1}$, and $10^0$, respectively.
        The lines show the fitting formula~\eqref{eq: binary_form} with the corresponding value of $\delta_\mathrm{pair}$ and $f_\mathrm{PBH}$.
    }
    \label{fig: bin_form_time}
\end{figure}

We have used the fitting formulas in Eqs.~\eqref{eq: binary_amp} and \eqref{eq: binary_form} for binary formation.
Here, we compare the fitting formulas with the numerical results.
Fig.~\ref{fig: bin_form_amp} shows the radial separation $\chi$ for the given initial density contrast $\delta_\mathrm{pair}$, and Fig.~\ref{fig: bin_form_time} shows the binary formation time for the given initial density contrast $\delta_\mathrm{pair}$.
We find that the fitting formulas are in good agreement with the numerical results for $\delta_\mathrm{pair}$ and $f_\mathrm{PBH}$ in the relevant parameter range.

\section{Merger rates with and without the binary formation condition}
\label{app: wo clust consis check}

\begin{figure}[t]
    \centering
    \includegraphics[width=.75\textwidth ]{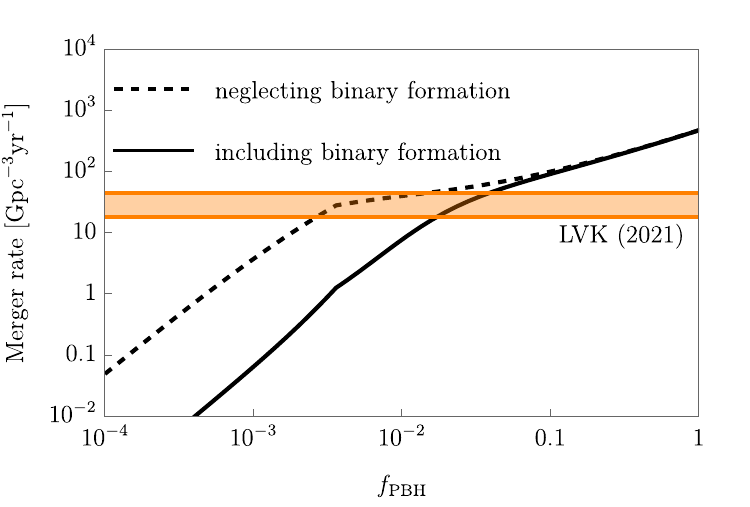}
    \caption{
        Reproduction of the merger rate of Poisson-distributed PBHs reported in Ref.~\cite{Raidal:2024bmm}.
        When the binary formation condition is neglected, following the calculation presented therein, the reported result is reproduced (dashed).
        When the binary formation condition is included, motivated by the physical picture discussed therein, the result shown by the solid line is obtained.
        The orange shaded band is the same as in Fig.~\ref{fig: result}.
    }
    \label{fig: wo clust}
\end{figure}

Here, we compare our treatment of the binary formation condition with that in the previous work~\cite{Raidal:2024bmm} focusing on the upper bound on the initial separation of binaries, $x_\mathrm{in}$.
To this end, we totally adopt the formalization in Ref.~\cite{Raidal:2024bmm} and then evaluate the merger rates with and without the upper bound on $x_\mathrm{in}$.
We show the merger rate for PBHs without clustering in Fig.~\ref{fig: wo clust}.
The dashed line shows the calculation following Ref.~\cite{Raidal:2024bmm}, which neglects the upper limit of $x_\mathrm{in}$ derived from the binary formation condition, in order to obtain the analytic formula of the merger rate.
This completely reproduces Fig.~8 of Ref.~\cite{Raidal:2024bmm} and shows that the LVK inference is achieved with $f_\mathrm{PBH}\sim 10^{-3}\,\text{--}\,10^{-2}$.
If we take into account the binary formation condition, i.e., including the upper limit of the integration over $x_\mathrm{in}$ in Eq.\,\eqref{eq: merger rate}, we obtain the merger rate shown by the solid line.
Here, we use the same binary and three-body formation conditions as Ref.~\cite{Raidal:2024bmm}.
We find that the merger rate is suppressed by including the condition for binary formation, particularly for small $f_\mathrm{PBH}$.
As a result, the value of $f_\mathrm{PBH}$ favored by the LVK merger rate is also shifted.
This shows that the upper bound on $x_\mathrm{in}$ cannot be neglected in evaluating the merger rate.

\section{Comparison with the suppression factor from Poisson distribution}
\label{app: consis check}

\begin{figure}[t]
    \centering
    \includegraphics[width=.75\textwidth ]{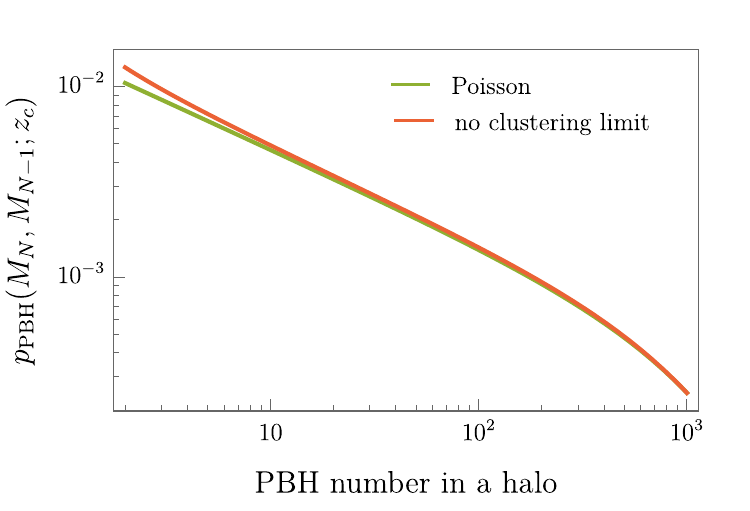}
    \caption{
        Comparison of $p_N$ for $f_\mathrm{PBH} = 1$ in the no-clustering case. The green and red lines denote $p_N$ calculated using the exact Poisson distribution and the no-clustering limit, assuming Gaussian density perturbations, respectively.
    }
    \label{fig: comparison pN}
\end{figure}

Here, we check the consistency between our evaluation of the suppression factor assuming the Gaussian distribution of PBHs and that derived using the Poisson distribution in Ref.~\cite{Raidal:2024bmm}.
To this end, we take the no-clustering limit.
We first consider the consistency of $p_\mathrm{PBH}$.
Fig.~\ref{fig: comparison pN} shows $p_\mathrm{PBH}(M_N,M_{N-1};z_c)$ as a function of $N$ for $f_\mathrm{PBH}=1$.
In the large $N$ limit, our result approaches that in the previous work.
In fact, the calculation of $p_N$ using the Press-Schechter formalism matches the Poisson statistical calculation in the large $N$ limit for large $f_\mathrm{PBH}$.
Even for $N = 2$, the relative difference is within 20\%.

\begin{figure}[t]
    \centering
    \includegraphics[width=.75\textwidth ]{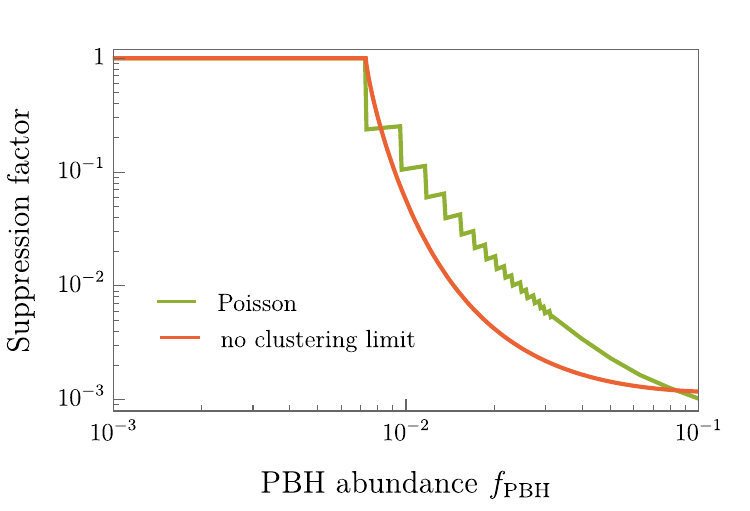}
    \caption{
        Comparison of the suppression factor for the exact Poisson distribution (green) and the no-clustering limit with Gaussian density perturbations (red). 
    }
    \label{fig: comparison SL}
\end{figure}

Figure~\ref{fig: comparison SL} shows the suppression factor in the no-clustering limit.
Here, we find that the suppression factor itself can differ by a factor of at most $5$.

\section{Comparison of angular momentum with previous work}
\label{app: comparison}

In the previous work~\cite{Kawasaki:2021zir}, they assumed that the angular momentum of a PBH binary is sourced by the gravitational interaction with the PBH nearest to the binary.
In this case, the angular momentum is given by 
\begin{align}
    j 
    =
    \left( \frac{x_\mathrm{in}}{Y} \right)^3
    \ ,
\end{align}
where we assumed a monochromatic mass function for PBHs, and $Y$ denotes the comoving distance between the binary and the third PBH.
Then, we can express the probability distribution for $j$ with that for $Y$.
The probability distribution for $Y$ is given by
\begin{align}
    \frac{\mathrm{d} P_Y(Y;y)}{\mathrm{d} Y}
    =
    4\pi Y^2 (1 + \xi(Y)) \bar{n}_\mathrm{PBH} 
    e^{-\Gamma(Y)+\Gamma(y)}
    \ ,
\end{align}
which is normalized for the integration over $y \leq Y < \infty$.

Then, the probability distribution $P(j,r_a|x_\mathrm{in},y)$ [see Eq.~\eqref{eq:prob_dist_binary_formation}] is written as 
\begin{align}
    \frac{\mathrm{d} P(j,r_a|x_\mathrm{in},y)}{\mathrm{d} j}
    &=
    \left| \frac{\mathrm{d} Y}{\mathrm{d} j} \right|
    \left. \frac{\mathrm{d} P_Y(Y;y)}{\mathrm{d} Y} \right|_{Y = j^{-1/3} x_\mathrm{in}}
    \nonumber \\
    &=
    \frac{x_\mathrm{in}}{3j^{4/3}} 
    \left. \frac{\mathrm{d} P_Y(Y;y)}{\mathrm{d} Y} \right|_{Y = j^{-1/3} x_\mathrm{in}}
    \ .
\end{align}
Using $Y$, we can rewrite Eq.~\eqref{eq: merger rate mid} as
\begin{align}
    R^{(y)}(t_0)
    &=
    \iint \mathrm{d} n_\mathrm{pair}(x_\mathrm{in},y) \mathrm{d}j
    \frac{\mathrm{d}P(j,r_a|x_\mathrm{in},y)}{\mathrm{d}j}
    \delta(t_0-\tau(r_a,j))
    \nonumber \\
    &=
    \iint \mathrm{d} n_\mathrm{pair}(x_\mathrm{in},y) \mathrm{d}Y \,
    \frac{\mathrm{d} P_Y(Y;y)}{\mathrm{d} Y}
    \delta(t_0-\tau(r_a,j(Y)))
    \nonumber \\
    &=
    \frac{1}{2} 
    \int_0^y \mathrm{d} x_\mathrm{in} \int_y^\infty \mathrm{d} Y \,
    4\pi x_\mathrm{in}^2 4 \pi Y^2 
    \bar{n}_\mathrm{PBH}^3
    (1 + \xi(x_\mathrm{in})) (1 + \xi(Y))
    e^{-\Gamma(Y)}
    \delta(t_0-\tau(r_a,j(Y)))
    \ ,
\end{align}
which matches the formula in the previous work~\cite{Kawasaki:2021zir} except for the factor $1/2$ to avoid the double counting of PBHs in binaries.

\begin{figure}[t]
    \centering
    \includegraphics[width=.73\textwidth ]{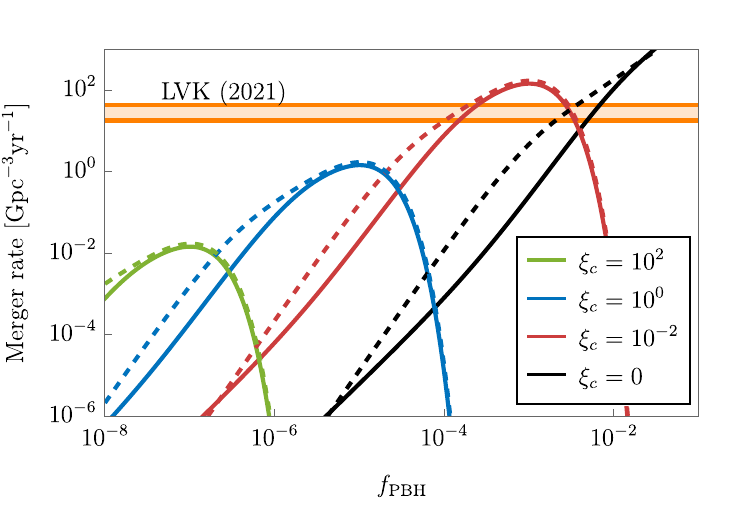}
    \caption{
        Merger rates of clustered PBHs for $\alpha=1.5$ by using the binary formation condition used in Ref.~\cite{Kawasaki:2021zir}.
        While the solid lines show the merger rates with the angular momentum contributed by all outer PBHs and matter density fluctuations, the dashed lines show those only with the angular momentum induced by the nearest outer PBH \protect\footnotemark .
        The black, red, blue, and green lines denote the different strengths of the correlation $\xi_c=0,~10^{-2},~10^0,~10^2$, respectively.
        The orange shaded band is the same as in Fig.~\ref{fig: result}.
        We do not include the suppression factor here.
        }
    \label{fig: strong clust}
\end{figure}

Figure~\ref{fig: strong clust} shows the merger rate for PBHs including the clustered cases.
For comparison purposes, we take the binary and three-body formation condition to be the same as in Ref.~\cite{Kawasaki:2021zir}, i.e., $\delta_\mathrm{pair} >1$ and $\delta_\mathrm{NN} >1$.
While the solid lines show the merger rates with the angular momentum contributed by all outer PBHs and matter density fluctuations, the dashed lines show those only with the angular momentum induced by the nearest outer PBH.
The merger rate is suppressed compared to Ref.\,\cite{Kawasaki:2021zir}.
This is because the additional contribution to the angular momentum makes the coalescence time longer, suppressing the merger rate observed at the current time.

\footnotetext{Equation (57) in Ref.~\cite{Kawasaki:2021zir} uses an incorrect numerical value, and the reported results are based on it. For this figure, we quote Ref.~\cite{Kawasaki:2021zir} for comparison, but recompute the results using the correct value for that equation.}

\bibliographystyle{JHEP}
\bibliography{Ref}

\end{document}